\documentclass[useAMS,usenatbib]{mn2e} 
\usepackage{times}  

\usepackage{rotating}
\usepackage[T1]{fontenc}
\usepackage{aas_macros}
\usepackage{amssymb}
\usepackage{amstext}
\usepackage{float}
\usepackage{graphicx}
\usepackage{caption}
\usepackage{subcaption}
\usepackage{color}
\usepackage{array}
\usepackage{changepage}
\usepackage{longtable}
\usepackage{amsmath} 
\usepackage{nth}

\DeclareRobustCommand{\ion}[2]{%
\relax\ifmmode
\ifx\testbx\f@series
{\mathbf{#1\,\mathsc{#2}}}\else
{\mathrm{#1\,\mathsc{#2}}}\fi
\else\textup{#1\,{\mdseries\textsc{#2}}}%
\fi}

\newcommand{\angstrom}{\textup{\AA}}

\providecommand{\e}[1]{\ensuremath{\times 10^{#1}}}

\newcommand{\Ha}{\rm H$\alpha$}
\newcommand{\Hb}{\rm H$\beta$}
\newcommand{\OHb}{[\ion{O}{iii}]\,$\lambda$5007 / H$\beta$}
\newcommand{\Oiii}{[\ion{O}{iii}]\,$\lambda$5007}
\newcommand{\NHa}{[\ion{N}{ii}]\,$\lambda$6583 / H$\alpha$}
\newcommand{\Nii}{[\ion{N}{ii}]\,$\lambda$6583}
\newcommand{\Moy}{${\rm M}_{\odot}$\,yr$^{-1}$}

\title[Can we trust aperture corrections?]{The SAMI Galaxy Survey: Can we trust aperture corrections to predict star formation?}
\author[Richards et al.]
{\parbox{\textwidth}{\raggedright
{S.~N.~Richards$^{1,2,3}$\thanks{E-mail:  \texttt{samuel@physics.usyd.edu.au}},
J.~J.~Bryant$^{1,2,3}$,
S.~M.~Croom$^{1,3}$, 
A.~M.~Hopkins$^{2}$,
A.~L.~Schaefer$^{1,2,3}$, 
J.~Bland-Hawthorn$^{1}$,
J.~T.~Allen$^{1,3}$,
S.~Brough$^{2}$,
G.~Cecil$^{4,1}$
L.~Cortese$^{5}$,
L.~M.~R.~Fogarty$^{1,3}$,
M.~L.~P.~Gunawardhana$^{6}$,
M.~Goodwin$^{2}$,
A.~W.~Green$^{2}$,
I.\,-T.~Ho$^{7,8}$,
L.~J.~Kewley$^{8}$,
I.~S.~Konstantopoulos$^{9,2}$,
J.~S.~Lawrence$^{2}$,
N.~P.~F.~Lorente$^{2}$,
A.~M.~Medling$^{8}$
M.~S.~Owers$^{2,10}$,
R.~Sharp$^{8,3}$,
S.~M.~Sweet$^{8}$,
E.~N.~Taylor$^{11}$}}\vspace{0.4cm} \\
\parbox{\textwidth}{$^{1}$Sydney Institute for Astronomy, School of Physics, University of Sydney, NSW 2006, Australia\\
$^{2}$Australian Astronomical Observatory, PO Box 915, North Ryde, NSW 1670, Australia\\
$^{3}$CAASTRO: ARC Centre of Excellence for All-sky Astrophysics\\
$^{4}$Department of Physics and Astronomy, University of North Carolina, Chapel Hill, NC 27510, USA\\
$^{5}$International Centre for Radio Astronomy Research, University of Western Australia, 35 Stirling Hwy, Crawley, WA 6009, Australia\\
$^{6}$Institute for Computational Cosmology and Centre for Extragalactic Astronomy, Department of Physics, Durham University, South Road, Durham, \hspace{1cm} DH1 3LE, UK\\
$^{7}$Institute for Astronomy, University of Hawaii, 2680 Woodlawn Drive, Honolulu, HI 96822, USA\\
$^{8}$Research School of Astronomy and Astrophysics, Australian National University, Cotter Rd., Weston, ACT 2611, Australia \\
$^{9}$Envizi, Level 2, National Innovation Centre, Australian Technology Park, 4 Cornwallis Street, Eveleigh NSW 2015, Australia \\
$^{10}$Department of Physics and Astronomy, Macquarie University, NSW 2109, Australia \\
$^{11}$School of Physics, The University of Melbourne, Parkville, VIC 3010, Australia}
\vspace{-0.4cm}}

\begin{document}

\date{Received, ** 2015, Accepted *** }

\pagerange{\pageref{firstpage}--\pageref{lastpage}} \pubyear{2015}

\maketitle

\label{firstpage}

\begin{abstract}
In the low redshift Universe ($z<0.3$), our view of galaxy evolution is primarily based on fibre optic spectroscopy surveys. Elaborate methods have been developed to address aperture effects when fixed aperture sizes only probe the inner regions for galaxies of ever decreasing redshift or increasing physical size. These aperture corrections rely on assumptions about the physical properties of galaxies. The adequacy of these aperture corrections can be tested with integral-field spectroscopic data. We use integral-field spectra drawn from $1212$ galaxies observed as part of the SAMI Galaxy Survey to investigate the validity of two aperture correction methods that attempt to estimate a galaxy's total instantaneous star formation rate. We show that biases arise when assuming that instantaneous star formation is traced by broadband imaging, and when the aperture correction is built only from spectra of the nuclear region of galaxies. These biases may be significant depending on the selection criteria of a survey sample. Understanding the sensitivities of these aperture corrections is essential for correct handling of systematic errors in galaxy evolution studies.
\end{abstract}

\begin{keywords}
galaxies: evolution -- techniques: spectroscopic \vspace{-0.4cm}
\end{keywords}


\section{Introduction}

Over the past decade, aperture correction methods have been developed to obtain global properties of galaxies by extrapolating measurements from a single spectrum probing only the central regions of each galaxy. When a nearby galaxy is spectroscopically observed with a single aperture, such as an optical fibre with a diameter on-sky of a few arcseconds, only the central region of a galaxy is typically probed for redshifts $z\lesssim0.3$. The magnitude of an aperture effect scales with both redshift and the physical size of a galaxy. 

\vspace{0.5cm}

\begin{figure}
\centering
\centerline{\includegraphics[width=\linewidth]{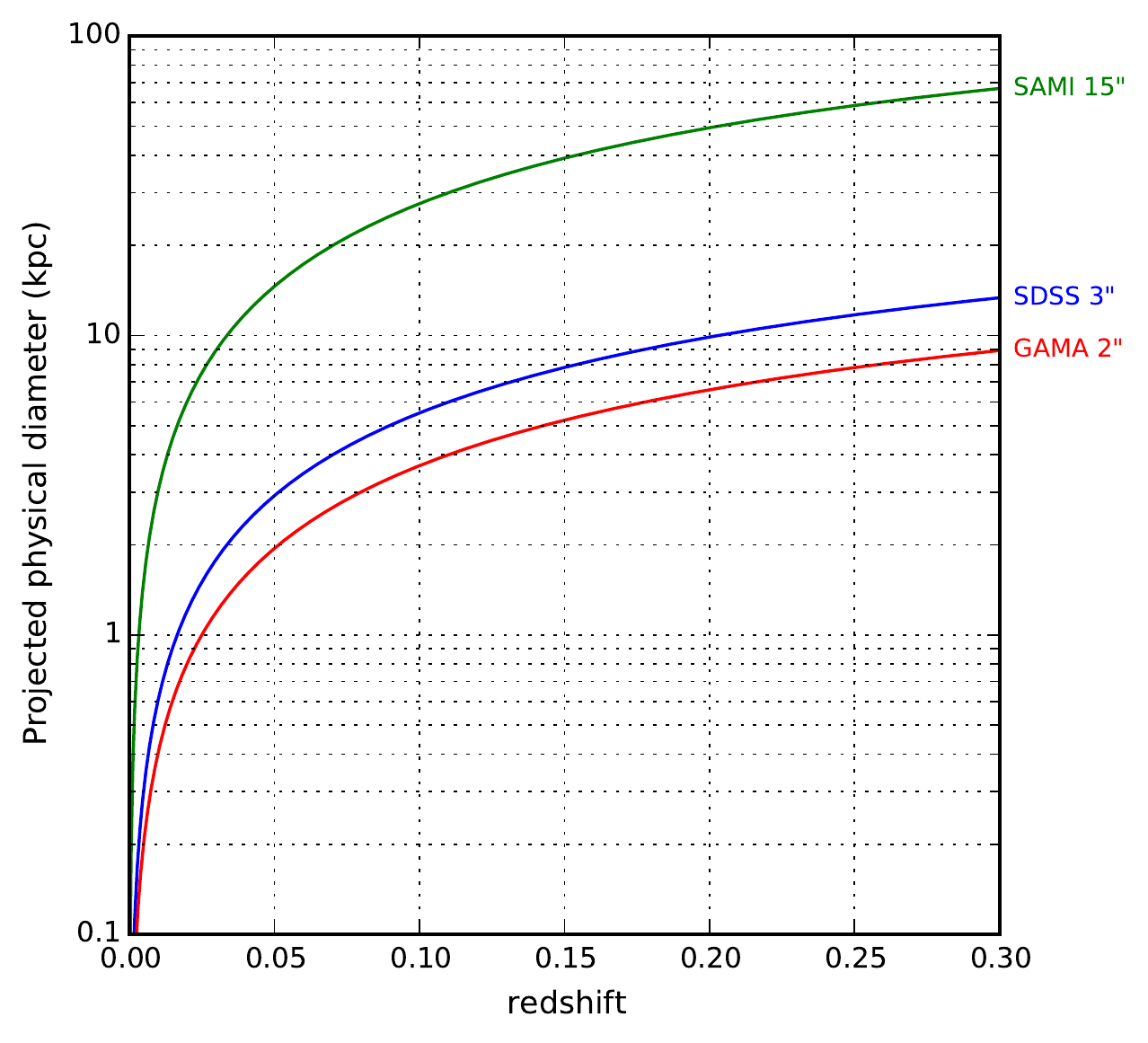}}
\caption{The projected physical sizes of the GAMA ($2$~arcsec, red), SDSS ($3$~arcsec, blue) and SAMI ($15$~arcsec, green) apertures as a function of redshift. For galaxies with redshift $z\lesssim0.2$, only the central few kpc are observed spectrally in GAMA and SDSS. \vspace{0.4cm}}
\label{fig:FvZ}
\end{figure}

The largest single aperture galaxy surveys to date are the Sloan Digital Sky Survey \citep[SDSS\footnote{http://www.sdss3.org/};][]{2000AJ....120.1579Y} and the Galaxy And Mass Assembly survey \citep[GAMA\footnote{http://www.gama-survey.org/};][]{2009A&G....50e..12D}. Both are optical spectroscopic surveys of $\approx10^{5}$~to~$10^{6}$ nearby galaxies with $z\lesssim0.3$, and have on-sky fibre diameters of $3$ and $2$~arcsec respectively. Therefore, the star formation rate (SFR) of galaxies within these surveys are subject to aperture effects. Figure \ref{fig:FvZ} shows the equivalent physical scale of an aperture's on-sky diameter as a function of redshift. By design however, GAMA incorporate spectra from other sources, including SDSS for bright galaxies. 

\vspace{0.4cm}

The SFR aperture correction used in GAMA and SDSS are different, with GAMA using a method prescribed by \citet[][hereafter H03]{2003ApJ...599..971H}, and SDSS that presented by \citet[][hereafter B04]{2004MNRAS.351.1151B}. For the benefit of the reader, a short summary of each method is provided.

\vspace{0.0cm}
\subsection{\citet{2003ApJ...599..971H} method (H03, GAMA)}
In a detailed look at SFR indicators from multi-wavelength data ($1.4$~GHz to $u$-band luminosities), H03 found that multiplying the stellar absorption corrected \Ha\ equivalent-width, EW(\Ha), from the fibre spectrum by the galaxy's $k$-corrected Petrosian $r$-band luminosity, and correcting for the Balmer decrement, gave a good approximation to the galaxy's total SFR, given by: 

\begin{flalign}
\label{eq:H03eq}
{\rm SFR\left(H03\right)} = & \hspace{0.1cm} \frac{{\rm EW(H\alpha)} \times 10^{-0.4\left(M_{r}-34.10\right)}}{\rm SFRF} \notag\\ & \hspace{1.2cm} \cdot \frac{3 \times 10^{18}}{[6564.61\left(1+z\right)]^2} \cdot \left(\frac{{\rm BD}}{2.86}\right)^{2.36} ,
\end{flalign}

\noindent where EW(\Ha) is the stellar absorption corrected \Ha\ flux divided by the median continuum level of the spectrum about the \Ha\ emission line (we perfom the absorption correction by subtracting fitted stellar templates via {\sc LZIFU}), $M_r$ is the absolute $k$-corrected $r$-band Petrosian magnitude of the galaxy including a correction for Galactic extinction, $z$ is the flow-corrected redshift of the galaxy, and BD is the Balmer decrement (stellar absorption corrected ratio of \Ha~/~\Hb\ emission line fluxes) assuming a fixed Case-B recombination value of $2.86$ \citep{2001PASP..113.1449C, 2003adu..book.....D} with a reddening slope of $2.36$ \citep{1989ApJ...345..245C} and the dust as a foreground screen averaged over the galaxy. SFRF is the "star formation rate factor" to convert to solar masses per year, e.~g.~$1.27$\e{34} W, as given by \citet{1998ApJ...498..541K} assuming a \citet{1955ApJ...121..161S} initial mass function (IMF). 

Only galaxies classified as star forming (SF) via the \citet{2003MNRAS.346.1055K} limit in \citet[][hereafter BPT]{1981PASP...93....5B} diagnostics were considered in this aperture correction. It assumes that the galaxy's EW(\Ha) and Balmer decrement profiles are constant across the galaxy. For the current work we interpret this to mean that for an EW(\Ha) measured using different aperture radii, to the limit of $2$ Petrosian radii (hereafter $R_{\rm 2P}$), the H03 SFR derived in those apertures will be constant.

Due to the straightforward approach of this aperture correction, H03 has been widely used in determining the SFR of galaxies in single aperture surveys. In the absence of large integral field surveys, no formal error analysis of the assumptions in H03 has been possible. Consequently, no errors on the SFRs are provided in GAMA~DR2 \citep{2015MNRAS.452.2087L}. What has been examined is how well the H03 SFRs compare with SFRs derived from other indicators \citep[][Wang et al. \emph{in prep}]{2003ApJ...599..971H,2014ApJ...782...90C}. The limit of these studies lies in how to interpret the random and systematic errors due to different indicators tracing different star formation time scales.

\vspace{-0.4cm}
\subsection{\citet{2004MNRAS.351.1151B} method (B04, SDSS)}
Younger, hotter stars (that contribute most of the star formation component of \Ha\ emission) are observed to have bluer optical colours, so B04 included a 3-colour dependance in their aperture correction (using the SDSS filters $g,r,i$ at a rest-frame of $z=0.1$). The best way to think about the B04 aperture correction is not as an aperture correction equation, but rather an aperture correction cube. The $x,y$ axes are the $g$-- $r, r$-- $i$ colours, and the $z$ axis a histogram (likelihood-distribution) of the SFR divided by the $i$-band luminosity (a proxy for specific-SFR) for each galaxy in a given $g$-- $r, r$-- $i$ cell. B04 constructed this aperture correction cube using spectra from high signal-to-noise (s/n) star forming galaxies in SDSS, with the $g,r,i$ colours and \Ha-based SFR coming from within the fibre. 

Using this aperture correction cube, it is then possible to find the likelihood distribution of SFR when only optical colours are known (independent of aperture size or shape). To calculate the B04 SFR: (a) measure the \Ha\ SFR from within the fibre; (b) subtract the $g,r,i$ fibre flux from the $g,r,i$ total galaxy flux to find the $g,r,i$ colours of the galaxy's light outside of the fibre aperture (annulus magnitudes); (c) locate the cell where the colours of the annulus magnitudes lie on the aperture correction cube's  $g$-- $r, r$-- $i$ grid; (d) multiply the likelihood distribution of the located cell by the annulus' $i$-band luminosity to find the SFR likelihood distribution of the annulus; (e) calculate the B04 SFR by adding together the SFR measured in the fibre and the median SFR from the SFR likelihood distribution of the annulus. It is worth clarifying that the B04 method of predicting the SFR of the annulus is independent from the aperture (fibre). 

There are three main assumptions in B04's original approach to calculating the SFR for SDSS galaxies. The first assumption relates to the method of calculating the fibre SFR for all galaxy types (defined as SF, low-s/n SF, AGN/Composites from BPT diagnostics). For SF galaxies, the emission lines, predominantly \Ha, were used to find the fibre SFR by fitting models to the spectra \citep{2001MNRAS.323..887C}. For other galaxy types, the fibre SFR was found by using a relationship of specific-SFR to D4000 (so as to not be biased by non star forming contributions to the emission lines). This relationship was constructed using spectra from SF galaxies. In a detailed look at this assumption, \citet{2007ApJS..173..267S} (using UV-based SFRs) found that non-SF galaxies followed different relationships depending on their BPT classification. As such, B04 revised this relationship in more recent editions of their SDSS SFRs. This particular revision only applies to classifications other than SF galaxies, though all B04 SFRs are adjusted{\footnote{http://wwwmpa.mpa-garching.mpg.de/SDSS/DR7/}} in later editions due to improvements in the {\small \sc SDSS} data reduction pipeline and model fitting the photometry of the outer regions of each galaxy.

The second assumption, which is more directly related to the aperture correction cube, is that optical colours are a good indicator of the \Ha\ specific-SFR. The most obvious possible discrepancy is that the stellar continuum and \Ha\ emission vary on two different timescales ($\approx100$ and $10$~Myr respectively). B04 assume the uncertainties are not systematic and provided them as percentile ranges of the likelihood-distributions. \citet{2007ApJS..173..267S} quote average 1$\sigma$ errors on B04 SFRs in the range of $0.29$~to~$0.54$~dex depending on the BPT classification. The aperture correction cube has a degeneracy between stellar age, metallicity and dust, which is assumed to broaden the likelihood distribution for any given $g$-- $r, r$-- $i$ cell.

The third assumption is that the 3-colour relationship with SFR in the nucleus of a galaxy is the same for that of the disk. Constructing the aperture correction cube with only nuclear spectra could lead to regions on the $g$-- $r, r$-- $i$ grid that are biased, in particular for galaxies with redshift, $z<0.1$, and so lead to systematic errors in SFR.

\vspace{-0.4cm}
\subsection{Previous tests of H03 and B04 SFRs}

Slit-scanning data from the Nearby Field Galaxy Survey \citep[NFGS;][]{2000ApJS..126..271J,2000ApJS..126..331J} were used by \citet{2005PASP..117..227K} to look at the biases of aperture effects on SFR, metallicity and reddening. They found that if a single aperture (fibre) could capture $>20\%$ of the galaxy's light, the systematic and random errors from the aperture effects would be minimised, but if $<20\%$ then the aperture effects are substantial. This $20\%$ boundary corresponds to redshifts of $0.04$ and $0.06$ for SDSS and GAMA respectively.

Data obtained via integral-field spectroscopy (IFS) is the preferred method for testing aperture corrections, due to the data being spatially resolved. It enables spectroscopic comparisons of the nuclear region, the disk, and the integrated light. Studies of galaxies observed via IFS that look at the effect of aperture corrections include \citet{2012MNRAS.420..197G}, \citet{2013A&A...553L...7I}, and \citet{2013MNRAS.435.2903B}. \citet{2013A&A...553L...7I} use the data of $104$ star forming galaxies from the Calar Alto Legacy Integral Field Area Survey \citep[CALIFA;][]{2012A&A...538A...8S} to measure the curves-of-growth of the \Ha\ flux, Balmer decrement and EW(\Ha), and empirically find aperture corrections as a function of \emph{$r_{a}$}/\emph{$R_{50}$}, where \emph{$r_{a}$} is the radius of a single aperture and \emph{$R_{50}$} the half-light radius (the radius containing $50\%$ of the Petrosian flux in the $r$-band). \citet{2012MNRAS.420..197G} compare the B04 aperture corrections with a sample of $24$ SF galaxies observed with VIMOS \citep{2003SPIE.4841.1670L}. They compared the ratio of the B04 aperture corrected SFR to the fibre SFR with the ratio of the total \Ha\ flux to the \Ha\ flux contained within a $3$~arcsec aperture on their data cubes. They find on average for their sample that the B04 correction underestimates the aperture correction factor by a factor $\approx2.5$ with a large scatter. \citet{2013MNRAS.435.2903B} directly compare H03 and B04 SFRs with SFRs from IFS data of $18$ galaxies that span a range of environments, as observed with {\sc SPIRAL} \citep{2006SPIE.6269E..0GS}. They find a mean ratio of $1.26\pm0.23$ and $1.34\pm0.17$ respective to H03 and B04.

Although several studies compare the H03 and B04 SFRs of galaxies with SFRs from total \Ha, all are limited by errors from either small-sample statistics or the inability to disentangle measurement and calibration biases \citep{2009PASP..121..937C}. 


Until recently, nearly all IFS data have been obtained with monolithic IFUs, meaning that the time taken to gather the data has been lengthy and the sample numbers small ($\lesssim100$, normally a few dozen). Efforts have now been made towards obtaining IFS data of $>10^{3}$ galaxies with multi-object IFS, improving an IFS survey speed by over an order of magnitude on monolithic IFUs. Such instruments include the Sydney-AAO Multi-object Integral-field spectrograph \citep[SAMI\footnote{SAMI: http://sami-survey.org/};][]{2012MNRAS.421..872C,2015MNRAS.447.2857B}, Mapping Nearby Galaxies at APO Instrument \citep[MaNGA\footnote{MaNGA: http://www.sdss3.org/future/manga.php};][]{2015ApJ...798....7B,2015AJ....149...77D} and the K-band Multi-Object Spectrograph \citep[KMOS\footnote{KMOS: http://www.eso.org/sci/facilities/develop/instruments/kmos.html};][]{2006NewAR..50..370S,2013Msngr.151...21S}. 

For this work we use data obtained as part of the SAMI Galaxy Survey \citep{2015MNRAS.446.1567A,2015MNRAS.447.2857B,2015MNRAS.446.1551S}, which already has reduced IFS data on $1212$ galaxies at the time of writing, with a survey target sample of $3400$ over three years. With a large initial sample, the SAMI Galaxy Survey makes for an ideal dataset to test the robustness of the H03 and B04 aperture corrections methods.

In Section 2 we detail the observations and data reduction of the SAMI Galaxy Survey, the sample selection and cuts applied to the SAMI Galaxy Survey data, and the ancillary data of the SAMI Galaxy Survey important to this work. In Section 3 we perform tests of the H03 and B04 aperture corrections both indirectly and directly. In Section 4 we discuss biases of the H03 and B04 methods and implications these might have on literature results. In Section 5 we conclude on the trustworthiness of the H03 and B04 aperture corrections and provide advice for future single aperture studies. Throughout this paper, "SF" is in reference to galaxies or spectra that lie below the \citet{2003MNRAS.346.1055K} star formation line on the log$_{10}$(\OHb) vs log$_{10}$(\NHa) BPT diagram. We assume the standard $\Lambda$CDM cosmology with $\Omega_{m}=0.3$, $\Omega_{\Lambda}=~0.7$ and $H_0$ = $70$\,km\,s$^{-1}$\,Mpc$^{-1}$.

\vspace{-0.7cm}
\section{Observations and data reduction}
The data used in this work were obtained with SAMI, which deploys $13$ hexabundles \citep{2011OExpr..19.2649B, 2014MNRAS.438..869B} over a $1$ degree field at the Prime Focus of the 3.9m Anglo-Australian Telescope. Each hexabundle consists of $61$ circularly packed optical fibres. The core size of each fibre is $1.6$~arcsec, giving each hexabundle a field of view of $15$~arcsec diameter. All $819$ fibres ($793$ object fibres and $26$ sky fibres) feed into the AAOmega spectrograph \citep{2006SPIE.6269E..0GS}. For SAMI observing, AAOmega is configured to a wavelength coverage of $370$~to~$570$\,nm with $R=1730$ in the blue arm, and $625$~to~$735$\,nm with $R=4500$ in the red arm. A seven point dither pattern achieves near-uniform spatial coverage \citep{2015MNRAS.446.1551S}, with $1800$\,s exposure time for each frame, totalling $3.5$\,h per field. 

As described in \citet{2015MNRAS.446.1567A}, in every field, twelve galaxies and a secondary standard star are observed. The secondary standard star is used to probe the conditions as observed by the entire instrument. The flux zero-point is obtained from primary standard stars observed in a single hexabundle during the same night for any given field of observation. The raw data from SAMI were reduced using the AAOmega data reduction pipeline, {\small 2dfDR}{\sc}\footnote{{\small 2dfDR}{\sc} is a public data reduction package managed by the Australian Astronomical Observatory, see http://www.aao.gov.au/science/software/2dfdr.}, followed by full alignment and flux calibration through the SAMI Data Reduction pipeline (see \citet{2015MNRAS.446.1551S} for a detailed explanation of this package). In addition to the reduction pipeline described by \citet{2015MNRAS.446.1567A} and \citet{2015MNRAS.446.1551S}, the individual frames are now scaled to account for variations in observing conditions. Absolute $g$-band flux calibration with respect to SDSS imaging across the survey is unity with a $9$\% scatter, found by taking the ratio of the summed flux within a $12$~arcsec diameter aperture centred on the galaxy in a $g$-band SAMI IFU image and the respective SDSS $g$-band image smoothed to the SAMI seeing. 

Emission-line maps (most notably \Hb, \Oiii, \Ha\ and \Nii) of all galaxies in the SAMI Galaxy Survey were produced using the IFU emission-line fitting package, {\small LZIFU}{\sc} (see \citet{2014MNRAS.444.3894H} for a detailed explanation of this package). {\small LZIFU}{\sc} utilises {\small pPXF}{\sc} \citep{2004PASP..116..138C} for stellar template fitting \citep[MILES templates, ][]{2011A&A...532A..95F} and the {\small MPFIT}{\sc} library \citep{2009ASPC..411..251M} for estimating emission line properties. It is possible to perform multi-component fitting to each emission line with {\small LZIFU}{\sc}, although for the purpose of this work we chose to only use the single-component Gaussian fits. 

\vspace{-0.4cm}
\subsection{Sample selection}
\begin{figure}
\centering
\centerline{\includegraphics[width=\linewidth]{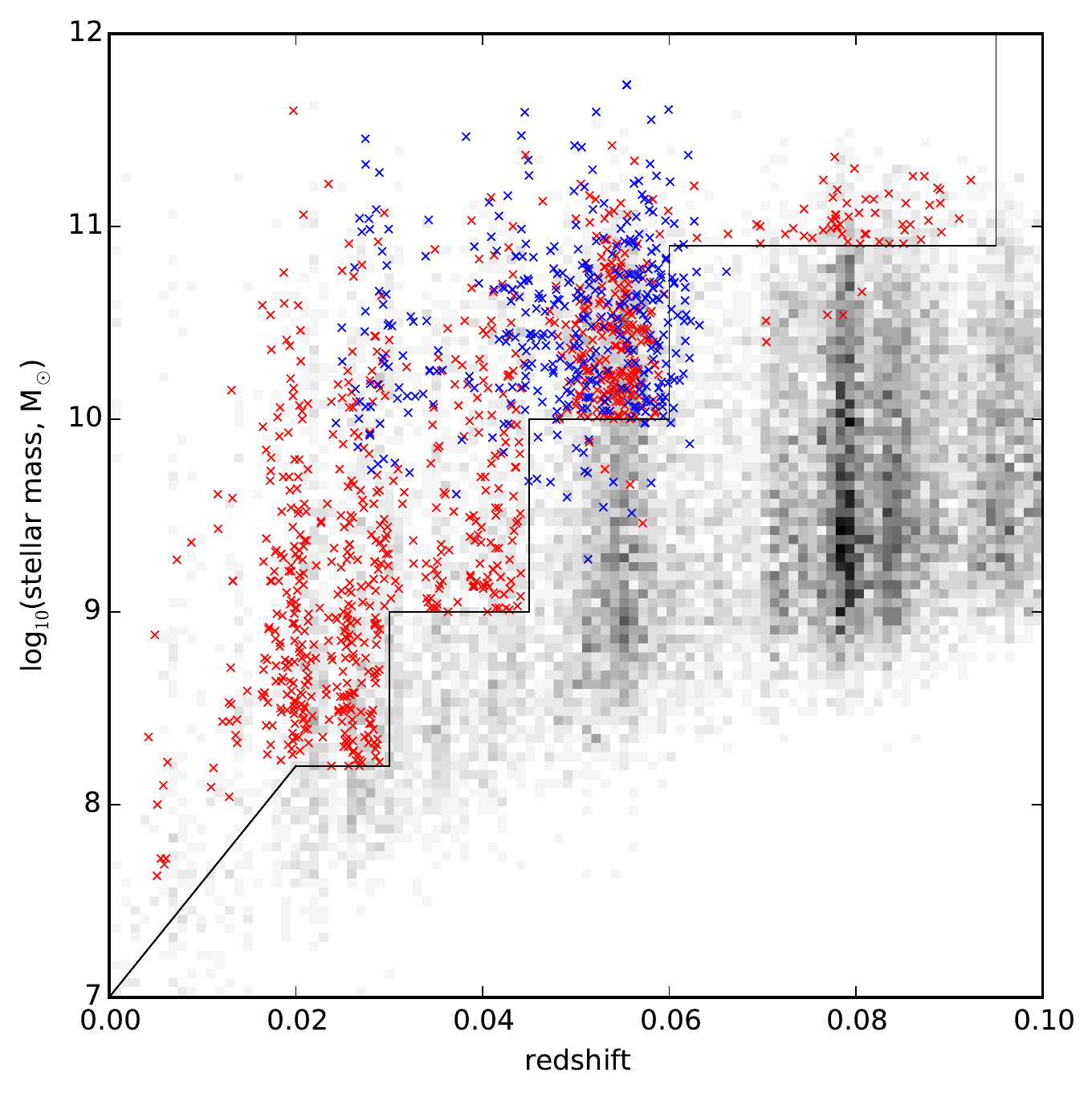}}
\caption{The location of our galaxies (red and blue points) overlaid on the SAMI Galaxy Survey target selection \citep[see Figure 4 of][]{2015MNRAS.447.2857B}. The red points are galaxies found in the GAMA regions, and the blue points those found in the Cluster regions. The background is a 2D histogram of the GAMA DR2 catalogue from which the SAMI field sample is drawn, with the black stepped-line representing the selection cut. Galaxies above this line are "Primary Targets". Galaxies that lie below this line are considered "Filler Targets" (included due to observational constraints on field tiling). Our sample of $1212$ galaxies is representative of the full SAMI target selection. \vspace{0.0cm}}
\label{fig:TS}
\end{figure}

At the time of writing, $1212$ galaxies had been observed as part of the SAMI Galaxy Survey (internal data release v$0.9$), and form the parent sample for the analysis in this work. The SAMI Galaxy Survey can be split into two populations of galaxies: those found in the GAMA regions (field galaxies) and those found in the Cluster regions (cluster galaxies). For a full description of the SAMI Galaxy Survey target selection, we refer the reader to \citet{2015MNRAS.447.2857B}. Of the $1212$ galaxies in our parent sample, $832$ are found in the GAMA regions and $380$ in the Cluster regions. Figure \ref{fig:TS} shows the stellar mass of these galaxies as a function of redshift, and reveals that the distribution of our parent sample is representative of the full SAMI Galaxy Survey's target selection. Different aspects of the analysis in this work use different subsamples of this parent sample, which are defined at the start of each section respectively.

\vspace{-0.4cm}
\subsection{Ancillary data}
The target selection of the SAMI Galaxy Survey \citep{2015MNRAS.447.2857B} allows for a plethora of existing multi-wavelength ancillary data, in particular for galaxies observed within the GAMA Survey fields ($2/3$ of the SAMI Galaxy Survey targets). Among many other properties, the GAMA~DR2 catalogue \citep{2015MNRAS.452.2087L} provides the Petrosian radii, stellar masses \citep{2011MNRAS.418.1587T}, S\'ersic fits with $R_e$ measurements \citep{2012MNRAS.421.1007K}, and spectroscopic redshifts and H03 aperture-corrected SFRs \citep{2013MNRAS.430.2047H,2013MNRAS.433.2764G} for every galaxy used in the analysis of this paper. Optical $u,g,r,i,z$ photometry is provided by SDSS DR10 \citep{2014ApJS..211...17A}, with the B04 total SFRs coming from the latest MPA-JHU Catalogue\footnote{B04 total SFRs are found in {"$\rm gal\_totsfr\_dr7\_v5\_2.fits.gz$"}, obtained at http://www.mpa-garching.mpg.de/SDSS/DR7/sfrs.html. The SFRs are derived from SDSS DR7 data \citep{2009ApJS..182..543A}, and include the corrections of \citet{2007ApJS..173..267S}.}. All stellar masses were found using the photometric prescription of \citet{2011MNRAS.418.1587T}. Following the scheme used by \citet{2014MNRAS.444.1647K} and \citet{2014ApJ...795L..37C}, visual morphological classification has been performed on the SDSS colour images by the SAMI Galaxy Survey team. Galaxies were divided into late- and early-types (or unclassified) according to their shape, presence of spiral arms and/or signs of star formation.

\vspace{-0.2cm}
\section{Testing of aperture corrections}
In this section we aim to provide analysis of the H03 and B04 aperture corrections using integral-field data from the SAMI Galaxy Survey. The analysis is divided into three sections, with the first being the comparison of SFRs from the H03 and B04 methods to that measured from SAMI galaxies, and the second and third being investigations into the assumptions of the H03 and B04 method respectively. 

\vspace{0.4cm}
\subsection{Comparing total SFRs from SAMI, GAMA \& SDSS}
The most common test of aperture corrections is in the comparison of the total \Ha\ SFRs measured from IFS data to that from an aperture correction. IFS data provides direct knowledge of the total instantaneous SFR of a galaxy (when full coverage is obtained), whereas the aperture corrections are predicting the total SFR indirectly. To do this comparison, from the $1212$ parent sample we selected galaxies that met the following criteria: (1) Matched to, and had measured SFRs in the GAMA and SDSS catalogues; (2) Classified as SF from the integrated SAMI spectrum via the \citet{2003MNRAS.346.1055K} limit in BPT diagnostics; (3) The SAMI hexabundle field of view probes out to at least $2$ effective radii ($2R_e$). After these cuts, we were left with $107$ galaxies. The SAMI SFRs were measured by binning the SAMI data cube, taking into account the spatial covariance \citep{2015MNRAS.446.1551S}, and fitting the binned spectrum with {\small LZIFU}{\sc}. The single-component emission line fits of the {\small LZIFU}{\sc} product were then used to compute the SFR via:

\begin{multline}
{\rm SFR\left(SAMI\right)} = \frac{{\rm H_{\alpha}} \cdot \left(4 \cdot \pi \cdot {\rm {d_{\emph{l}}}^{2}}\right)}{\rm SFRF} \cdot \left(\frac{{\rm BD}}{2.86}\right)^{2.36} ,
\label{eq:SAMI_SFR}
\end{multline}

\noindent where \Ha\ is the integrated flux (in ${\rm Wm^{-2}}$) of the single component Gaussian fit of the \Ha\ emission line after stellar continuum subtraction; SFRF is the star formation rate factor to convert to solar masses per year = $1.27$\e{34} W, as given by \citet{1998ApJ...498..541K} assuming a \citet{1955ApJ...121..161S} initial mass function (IMF) and solar metallicity; d$_{l}$ is the luminosity distance in meters; BD is the Balmer decrement (as described in Equation \ref{eq:H03eq} along with the reddening equation). We also ensure both H03 and B04 SFRs are scaled accordingly to match our use of a \citet{1955ApJ...121..161S} IMF.

\begin{figure}
\centering
\centerline{\includegraphics[width=7.9cm]{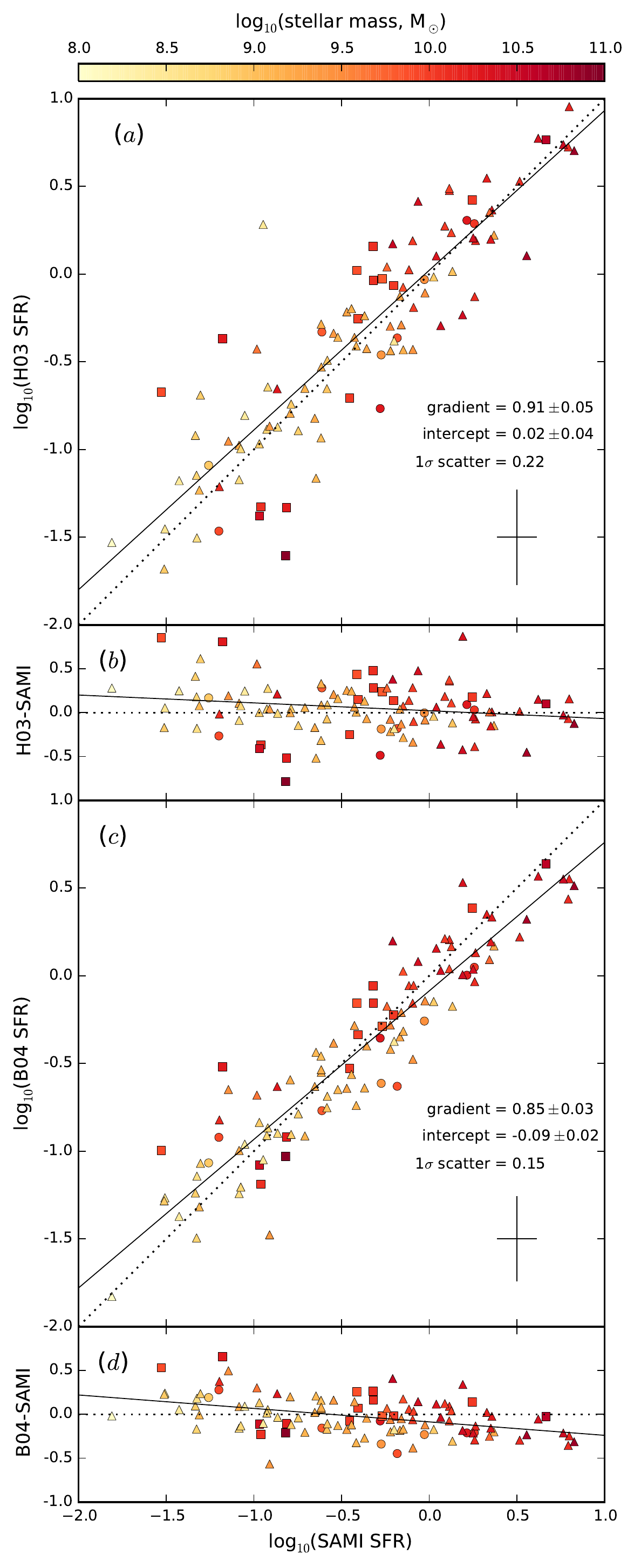}}
\caption{log$_{10}$(SFR) in \Moy\ of star forming galaxies found by the methods ($a$) of \citet[][H03]{2003ApJ...599..971H} and SAMI, and ($c$) \citet[][B04]{2004MNRAS.351.1151B} and SAMI. ($b,d$) shows the residuals from a $1$:$1$ correlation in ($a,c$) respectively. There are the same $107$ data points (galaxies) in all diagrams, with their colours representing the log$_{10}$(stellar mass, ${\rm M}_{\odot}$). Square, triangle and circle markers represent early-type, late-type and unclassified morphologies respectively. The typical error bars for these data are given in the lower right of ($a,c$). No formal error for the H03 SFR is given GAMA~DR2, so a typical error of the H03 method was taken from \citet{2003ApJ...599..971H}. The dotted lines are lines of the unity relation. The solid lines are least-squares fits to these data with the gradient, intercept and $1\sigma$ scatter about the fit shown in the lower right of ($a,c$). These data span approximately 4 orders of magnitude in SFR from $0.01$~to~$10$ \Moy.}
\label{fig:SAMI_H03_B04}
\end{figure}

Figure \ref{fig:SAMI_H03_B04} shows the comparison between the SAMI SFRs and SFRs from H03 and B04, and suggests there are slight trends with respect to the SAMI values in the H03 and B04 methods potentially biasing literature results that rely on them. Assuming the SAMI SFR to be the true SFR, the H03 method shows only over-estimation for galaxies with a low SFR, whereas the B04 shows both over- and under-estimation for low and high SFR galaxies respectively. H03 exhibits a larger scatter than B04 with scatters of $0.22$ and $0.15$~dex respectively. The best fits to these data for each aperture correction are given as:

\vspace{-0.2cm}

\begin{multline}
{\rm SFR\left(SAMI\right)} = \frac{{\rm SFR\left(H03\right)} - \left(0.02\pm0.04\right)}{\rm \left(0.91\pm0.05\right)} ,
\label{eq:H03_fit}
\end{multline}

\begin{multline}
{\rm SFR\left(SAMI\right)} = \frac{{\rm SFR\left(B04\right)} + \left(0.09\pm0.02\right)}{\rm \left(0.85\pm0.03\right)} ,
\label{eq:B04_fit}
\end{multline}

\noindent where all SFRs are in log$_{10}$(\Moy). After visually noticing a gradient in stellar mass in Figure \ref{fig:SAMI_H03_B04}($d$), we found no significant gain when including a stellar mass term for the B04 fit.

Comparing SFRs can reveal the presence of systematic errors, but as with all analysis of aperture corrections performed with this technique it is not possible to locate the origin of such errors from the SFRs alone. To locate biases in aperture corrections, rather than comparing SFRs from different methods, tests should be performed on the assumptions that go into the aperture corrections.

\begin{figure*}
\centering
\centerline{\includegraphics[width=14.2cm]{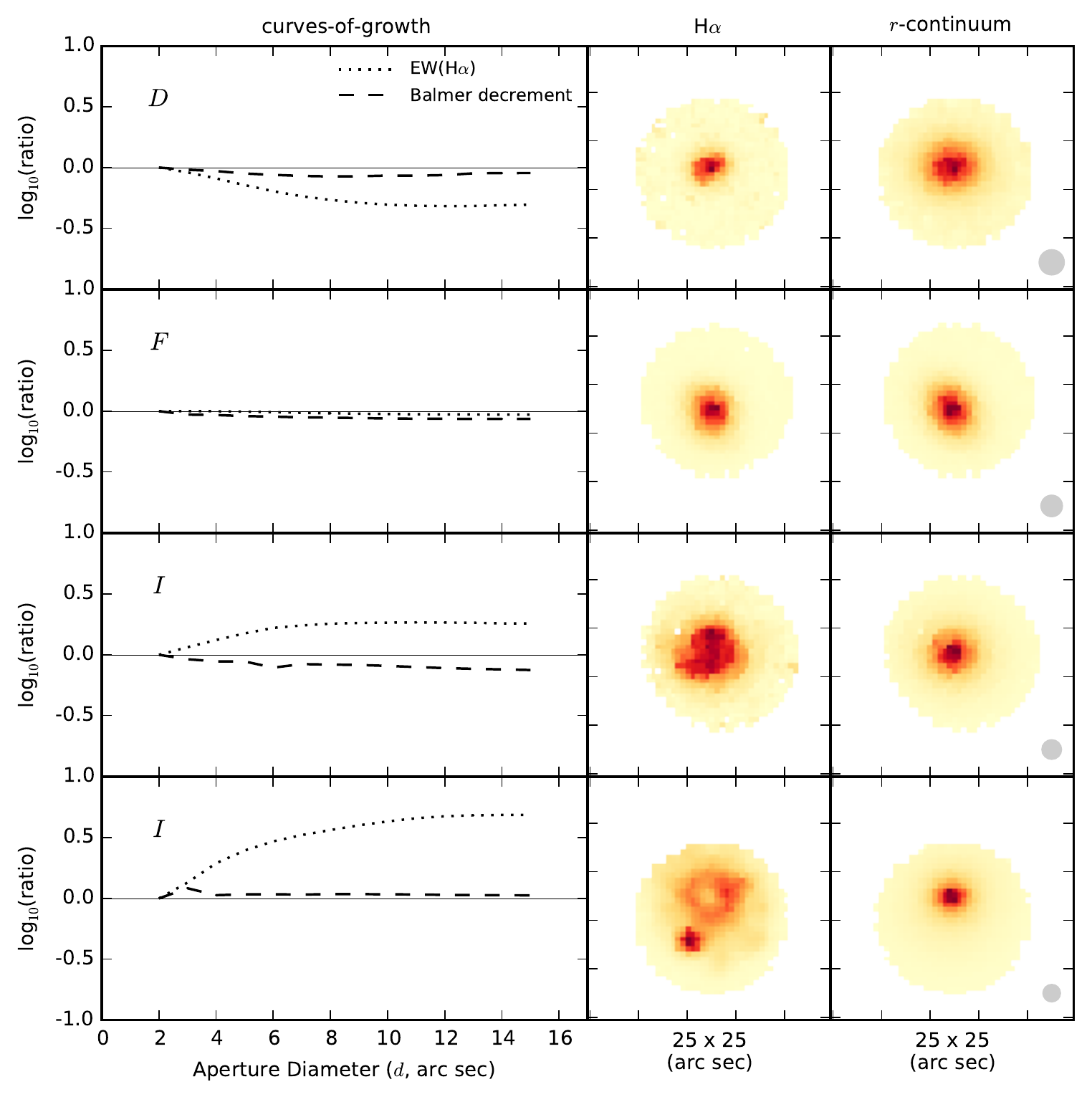}}
\caption{Example galaxies that fall within the respective curve-of-growth classifications: decreasing ($D$), flat ($F$) or increasing ($I$), where each row is a different galaxy. The first column shows the EW(\Ha) and Balmer decrement curves-of-growth (dotted and dashed line respectively). The curves-of-growth have been normalised to the measurement obtained with an aperture diameter of $2$~arcsec ("ratio"). The second and third columns are the SAMI \Ha\ and $r$-continuum maps for each galaxy (normalised to the maximum of each map for visual aid), and the size of the $g$-band PSF is given as a grey circle in the lower right of the $r$-continuum maps. All maps are $25\times25$~arcsec in size, and are orientated such that North is up and East is left. The Balmer decrement curves-of-growth tend to remain flat for all aperture sizes, but the EW(\Ha) varies greatly depending on the relative distributions of \Ha\ to $r$-continuum. \vspace{0.0cm}}
\label{fig:H03_EW_BD}
\end{figure*}

\vspace{-0.2cm}
\subsection{Testing the H03 aperture correction with SAMI data}
With IFS data it is possible to directly test the assumptions of H03 that a galaxy's EW(\Ha) and Balmer decrement profiles are flat. The form of Equation \ref{eq:H03eq} means that for a galaxy observed with ever increasing aperture sizes, the H03 SFR derived from the measured EW(\Ha) and Balmer decrement in those apertures should remain constant. If the EW(\Ha) and Balmer decrement profiles vary across a galaxy, the H03 SFR equated at ever-increasing aperture sizes should approach to the true total SFR when the aperture radius is equal to $2$~$\times$ the galaxy's $r$-band Petrosian radius ($R_{\rm 2P}$). 

Equation \ref{eq:H03eq} relies on three spectral measures: redshift, EW(\Ha) and Balmer decrement. The first step in this test was to see how the latter two vary for apertures of $d$ from $2$ to $15$~arcsec with a step of $1$~arcsec for all SF galaxies in the $1212$ parent sample that had an \Ha\ s/n $> 3$ for all apertures (leaving $461$ galaxies). Galaxies that were excluded due to this cut had either AGN/LINER emission or no reliable \Ha\ flux measurement in the smallest apertures. All galaxies that exhibited extra-nuclear star formation still had detectable \Ha\ flux in the smallest apertures. The spectrum for each aperture was obtained by binning all spaxels of the SAMI data cube (taking into account spatial covariance) that fell within the aperture footprint centred on the galaxy. The EW(\Ha) and Balmer decrement for each spectrum were found by fitting each spectrum with {\small LZIFU}{\sc}. The data for each galaxy were then normalised by its respective measurement at $d=2$, resulting in a curve-of-growth of each galaxy's EW(\Ha) and Balmer decrement (see leftmost column of Figure \ref{fig:H03_EW_BD}). 

Overall, the Balmer decrement curves-of-growth tend to be flat for all galaxy types (staying within a range of $0.1$~dex), but the EW(\Ha) curves-of-growth vary greatly (in extreme cases there can be more than an order of magnitude difference between $d=2$ and $d=15$). The EW(\Ha) curves-of-growth can be categorised as either decreasing ($155$ galaxies), flat ($149$ galaxies) or increasing ($157$ galaxies). The classifications were performed by allowing the flat ($F$) curves-of-growth to have a range of $\pm0.05$~dex between $d=2$ and $d=15$. Higher and lower than this range, the curves-of-growth were classified as increasing ($I$) and decreasing ($D$) respectively. 

Figure \ref{fig:H03_EW_BD} provides example galaxies for each classification, and it immediately becomes evident as to why the EW(\Ha) curves-of-growth vary so much when inspecting the \Ha\ and $r$-continuum maps (middle and rightmost columns). $D$ have more centrally concentrated \Ha\ compared to their $r$-continuum, $F$ have similar \Ha\ and $r$-continuum profiles, and $I$ fall into two subcategories: either the $r$-continuum shows a steeper radial decrease than the \Ha\ emission, or there are off-centred star forming regions (bright in \Ha) that don't show up in the $r$-continuum. The H03 aperture correction (Equation \ref{eq:H03eq}) relies on the EW(\Ha) and Balmer decrement curves-of-growth being flat, which this analysis shows is only true $1/3$ of the time.

\begin{figure}
\centering
\centerline{\includegraphics[width=\linewidth]{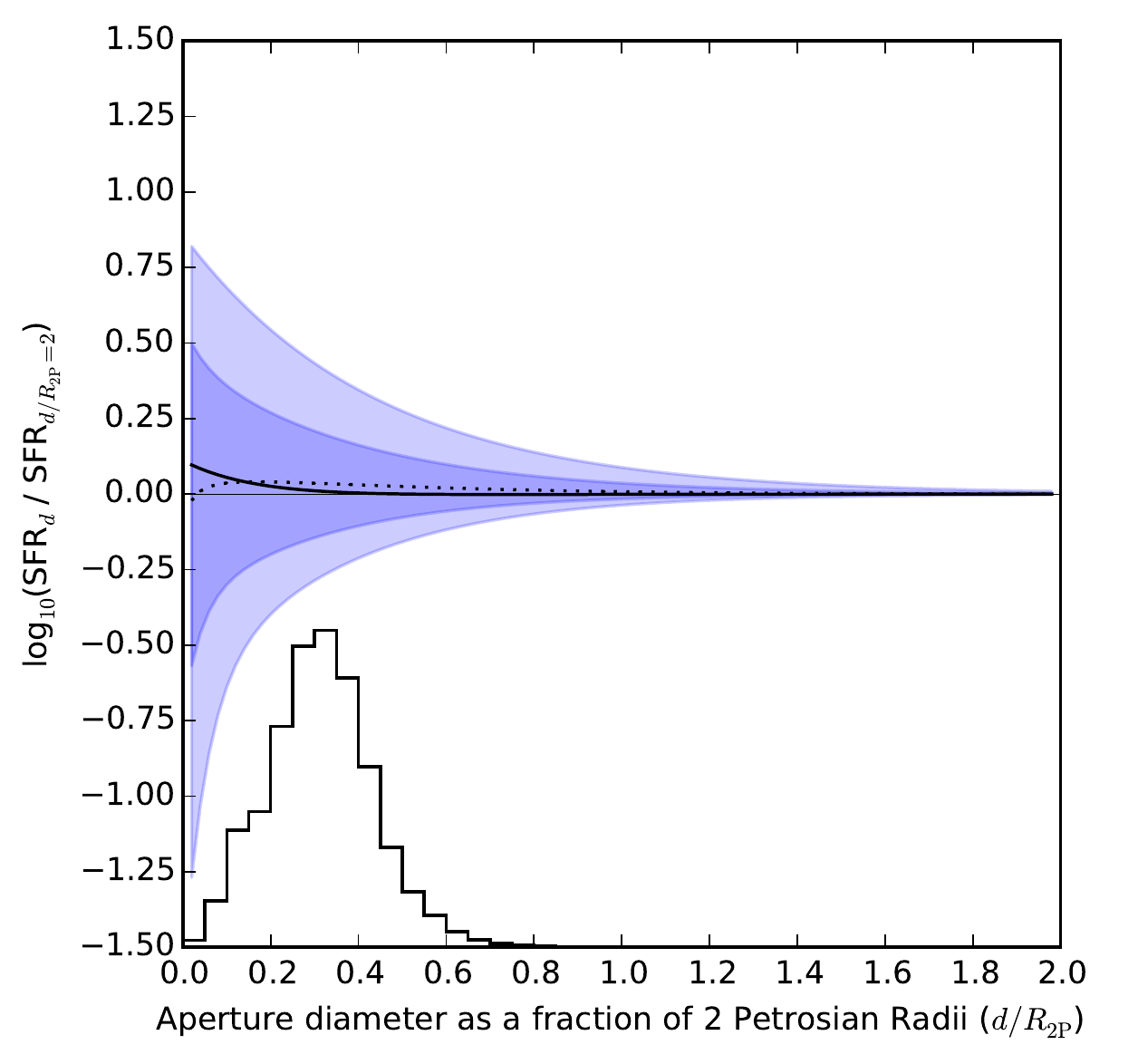}}
\caption{The error distribution of an H03 derived SFR as a function of aperture size. The percentile ranges for all $461$ H03 curves-of-growth are shown as shaded regions, and their curve-of-growth lines from the bottom-up are $2.5\%$, $16\%$, $50\%$ (median; thick line), $84\%$ and $97.5\%$. The dotted line is the curve-of-growth of the mean. The thin horizontal line is unity. The $y$-axis is log$_{10}$(SFR$_{d}$ / SFR$_{d/R_{\rm 2P}=2}$). The equations and coefficients of the fits to the percentiles can be found in Table \ref{tab:H03_cogs_fits}. The stepped-histogram shows the distribution of respective aperture sizes for all GAMA DR2 galaxies with redshift $z~<~0.1$, only including those measured with a $2$~arcsec aperture. This means that for a galaxy whose $d/R_{\rm 2P}=0.3$, the $1\sigma$-error on its H03 SFR is $0.18$~dex. For the smallest aperture sizes (i.e. large, nearby galaxies), the $1\sigma$-error becomes $\sim0.5$~dex, and the median departs from unity to become $\sim0.1$~dex, meaning H03 is more likely to over-estimate the SFR by $\sim0.1$~dex. This error is only due to aperture effects. To get the full uncertainty of SFR, random and systematic errors on the flux, modelling, initial mass function, etc would need be been taken into account. \vspace{1cm}}
\label{fig:H03_cogs}
\end{figure}

\begin{table}
\caption{Table of coefficients to find the errors for a galaxy's SFR after it has undergone the H03 aperture correction (see Figure \ref{fig:H03_cogs} for a description of these fits). Equation \ref{eq:H03_cog_eq} is to be used for calculating the percentiles. The "resid" column shows the median residual of the fit in dex for the range $0.01 < d/R_{\rm 2P} < 1.00$. The median ($50^{th}$-percentile) can be considered as an adjustment to the H03 SFRs. We do not presume to know the significance of the fitting coefficients to five decimal places, but they are provided for the sake of computation. \vspace{0.3cm}}  
\centering
\begin{tabular}{@{}r@{\hspace{0.3cm}}r@{\hspace{0.3cm}}r@{\hspace{0.3cm}}r@{\hspace{0.3cm}}r@{\hspace{0.3cm}}r@{}}
\noalign{\smallskip}
\hline
\noalign{\smallskip}
Percentile & A & B & C & D & resid \\ 
\hline
\noalign{\smallskip}
2.5  & -0.926(18) & -19.637(70) & -0.682(22) &  -2.919(96) & 0.013 \\
16.0 & -0.369(69) &  -3.129(41) & -0.365(93) & -25.097(86) & 0.004 \\
50.0 & -5.606(92) &  -4.807(18) &  5.717(52) &  -4.844(72) & 0.003 \\
84.0 &  0.442(29) &  -2.503(26) &  0.118(49) & -20.431(52) & 0.005 \\
97.5 &  0.432(06) &  -2.274(50) &  0.426(35) &  -2.274(49) & 0.014 \\
\hline
\end{tabular}\\
\label{tab:H03_cogs_fits}
\vspace{0.4cm}
\end{table}

To quantify the error of this assumption, we find the H03 SFR curve-of-growth for each galaxy over the same aperture range, and fit each curve-of-growth with an exponential, constrained such that the exponential has to reach within $1\%$ of its asymptote at $d/R_{\rm 2P}=2$ (by definition of Equation \ref{eq:H03eq}). A H03 SFR curve-of-growth can be fit with an exponential such that the residual on the fit is typically less than $0.05$~dex for all apertures. To put all the H03 curves-of-growth on the same diagram, we normalised each fit by the SFR found at $d/R_{\rm 2P}=2$, and converted the aperture sizes to units of $d/R_{\rm 2P}$. Combining the curves-of-growth like this enables us to measure the percentile ranges for different aperture sizes. A diagram of these percentiles can be found in Figure \ref{fig:H03_cogs}, which includes the H03 curves-of-growth from $461$ galaxies. The shape of the percentiles can be fitted with the analytical expression:

\begin{figure}
\centering
\centerline{\includegraphics[width=\linewidth]{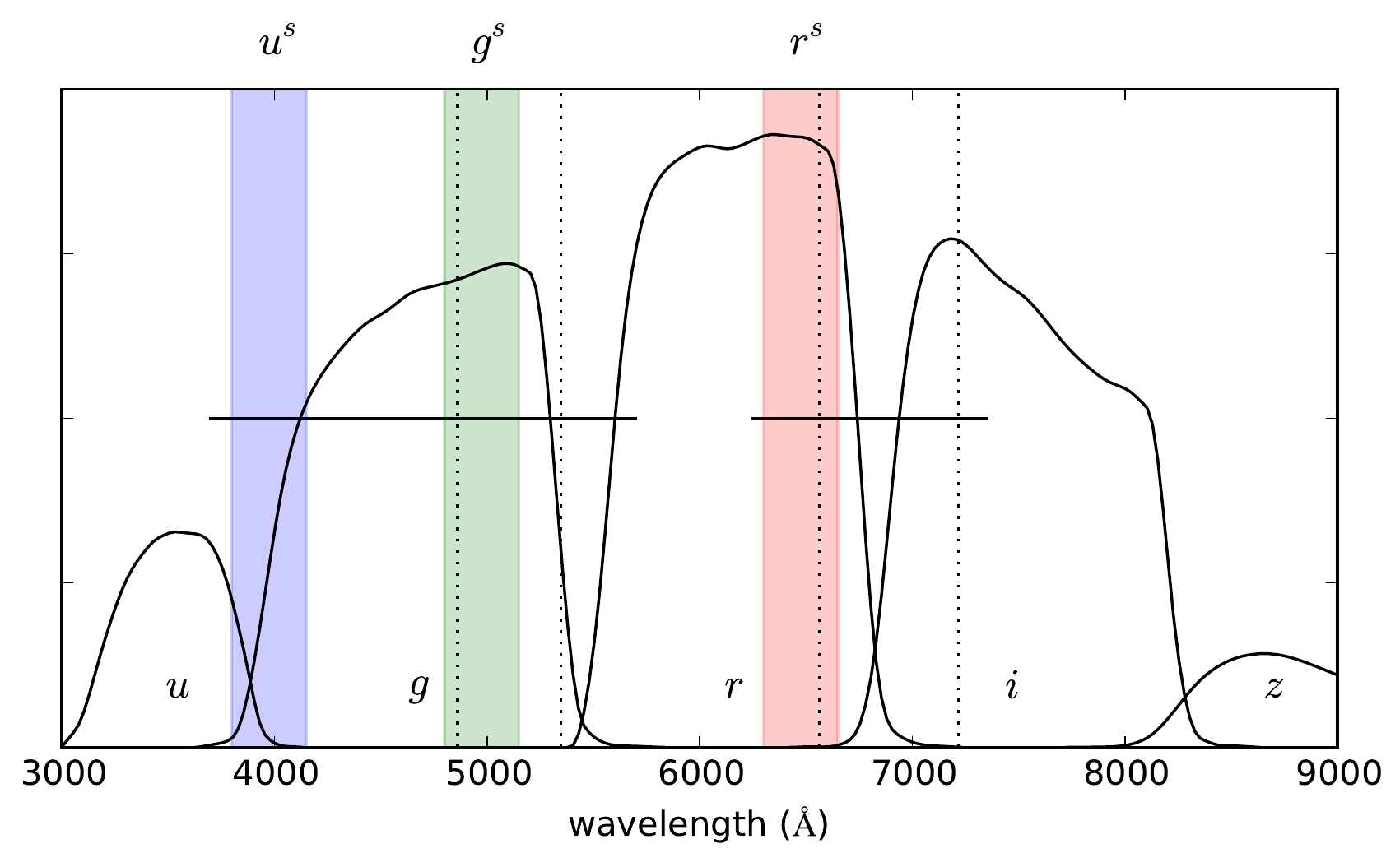}}
\caption{The custom filter set used to create the SAMI version of the aperture correction cube ($ACC$). The horizontal lines show the wavelength range of SAMI (Blue and Red arms of the AAOmega spectrograph). The vertical dotted lines show the limits of the locations of \Hb\ and \Ha\ for $0<z<0.1$. The blue, green and red shaded areas represent the wavelength coverage of the custom SAMI filter-set, labelled as $u^s,g^s,r^s$ respectively. The wavelength ranges of each filter are: $3800\angstrom<u^s<4150\angstrom$, $4800\angstrom<g^s<5150\angstrom$, $6300\angstrom<r^s<6650\angstrom$. The SDSS $u,g,r,i,z$ filters are also overlaid for comparison. The SAMI filter set was required because the SAMI data does not span the full SDSS $g,r,i$ filter range, due to the Red arm of SAMI having over double the spectral resolution of the Blue arm. \vspace{0.4cm}}
\label{fig:SAMI_filters}
\end{figure}

\begin{multline}
{\rm log}_{10}\left(error\right) = {\rm A}\cdot{\rm exp}\left({\rm B}\cdot \frac{d}{R_{\rm 2P}}\right) + {\rm C}\cdot{\rm exp}\left({\rm D}\cdot \frac{d}{R_{\rm 2P}}\right) ,
\label{eq:H03_cog_eq}
\end{multline}

\noindent where $error$ is the percentile error on the H03 aperture corrected SFR; A, B, C \& D are the coefficients given in \ref{tab:H03_cogs_fits} respective to their percentile; $d$ is the size of the aperture diameter in arcsec; $R_{\rm 2P}$ is the $2\times$ $r$-band Petrosian radius of the galaxy in arcsec. This analytical expression can be used to find the percentile error distribution on the H03 SFR for any given galaxy. For redshifts $z<0.1$, a GAMA~DR2 galaxy has a median $d/R_{\rm 2P}\approx0.3$, resulting in a $1\sigma$~error on its H03 SFR of $0.18$~dex. This error is only the error on the assumptions that go into the H03 aperture correction, and to get formal errors, the EW(\Ha) and Balmer decrement measurement errors would have to be included. We found no correlation between a galaxy's H03 SFR curve-of-growth and a global property of the galaxy (including: SFR at $d/R_{\rm 2P}=2$, stellar mass, $r$-band S\'ersic Index, Petrosian $g$-- $r$ colour, redshift and $5^{th}$ Nearest Neighbour environment density). 

\begin{figure}
\centering
\centerline{\includegraphics[width=9cm]{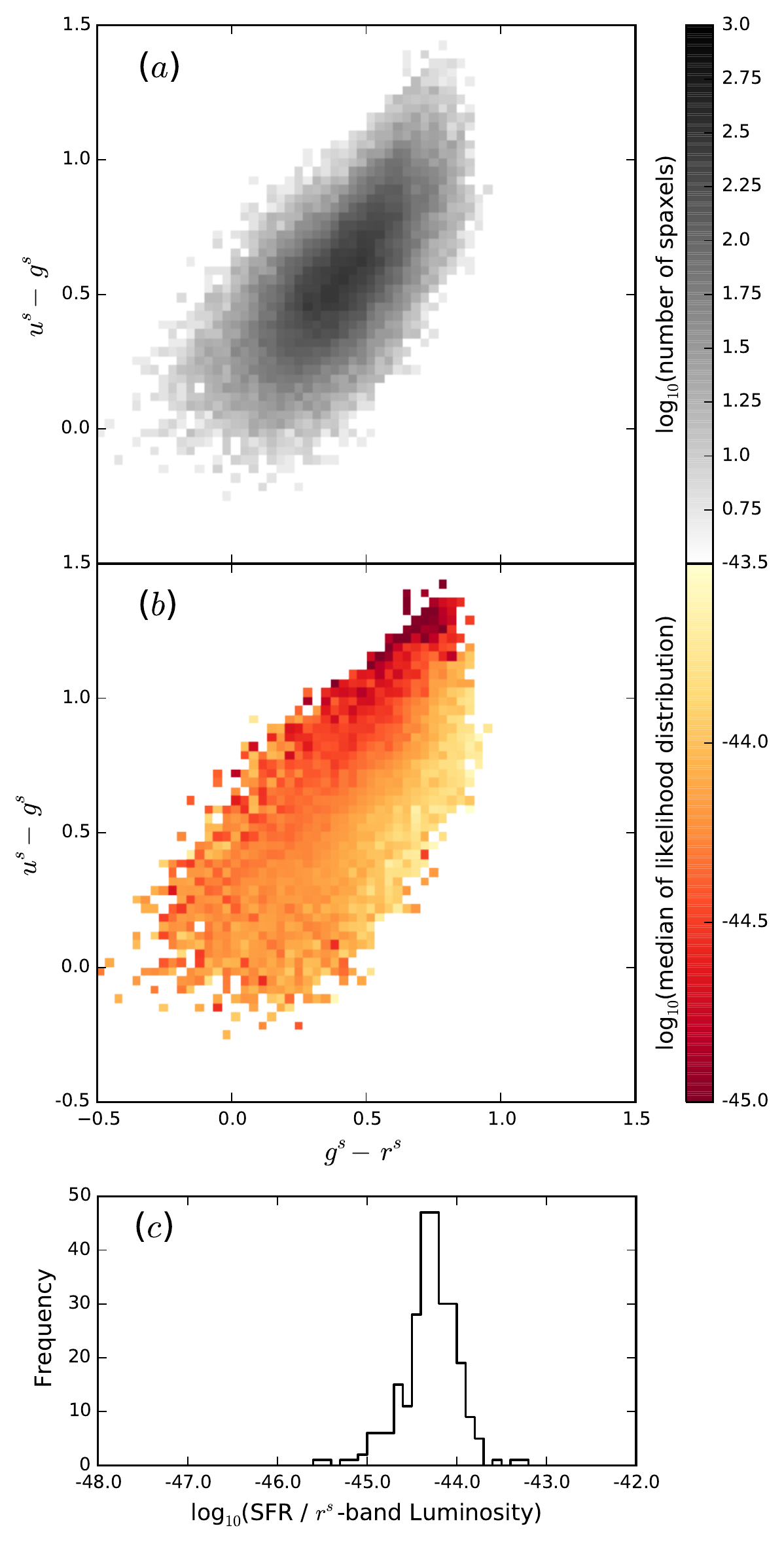}}
\caption{The SAMI version of the aperture correction cube ($ACC$), built from $48273$ spaxels. ($a$) shows the grid of $u^s$-- $g^s$ vs. $g^s$-- $r^s$, with the intensity being the log$_{10}$(number of spaxels) that contribute to each cell. Each cell is $0.04$~mag square in size. ($b$) is the same as ($a$), but the intensity is the log$_{10}$(median of the SFR/$r^s$-band luminosity likelihood distribution) for each cell. ($c$) is an example of the likelihood-distribution at the nominal point where $u^s$-- $g^s=0.5$, $g^s$-- $r^s=0.5$. $322$ spaxels contribute to this likelihood distribution, which has a median of $-44.3$ and a $1\sigma$ error of $0.35$~dex. \vspace{0.4cm}}
\label{fig:B04_cubes}
\end{figure}

\vspace{0.8cm}
\subsection{Testing the B04 aperture correction with SAMI data}
There are two assumptions that go into the B04 aperture correction cube that we can examine: (1) Broadband optical colours can act as a tracer of the \Ha-based SFR; (2) An aperture correction cube created from spectra probing only the nuclear regions of galaxies can be representative of a galaxy's disk. The widths of the SFR likelihood distributions that come from the aperture correction cube are representative of the errors due to the first assumption. B04 provide the percentiles of the SFR likelihood distributions of each galaxy. Obtaining a formal error on the second assumption from this analysis is not possible due to mismatching of available data between SAMI and B04, which will become clear as the analysis progresses.

To examine the assumption that broadband optical colours can act as a tracer of the SFR(\Ha), we first need to construct a SAMI version of the aperture correction cube (hereafter $ACC$). In B04, the SDSS optical filters $g,r,i$ are used to construct their aperture correction cube, but the wavelength range of SAMI does not span that entire filter set. Instead, we opt to use a custom top-hat filter set that can be applied to the spectra ($k$-corrected to $z=0$), taking the notation $u^s,g^s,r^s$ as they most closely match the standard $u,g,r$ filters respectively (see Figure \ref{fig:SAMI_filters}). The adoption of a custom filter set means that the magnitude of any bias or relation found with our data is not representative of the B04 aperture correction cube. The presence of a bias or relation, however, would indicate that one would likely also be present in the B04 aperture correction cube.

The native spaxel (spatial pixel) size of the SAMI data cubes is $0.5$~arcsec square, though to improve s/n, especially in the outer disks of galaxies, we opted to bin the data such that the spaxel size is now $1$~arcsec square. Taking all SF spaxels, we computed their $u^s,g^s,r^s$ magnitude colours, $r^s$-luminosity (in Watts) and SFR(\Ha) (Equation \ref{eq:SAMI_SFR}), only accepting spaxels with SFR~s/n~$>2$ (leaving $48273$ spaxels in total). These data formed the $ACC$ (see Figure \ref{fig:B04_cubes}). 

\begin{figure*}
\centering
\centerline{\includegraphics[width=16cm]{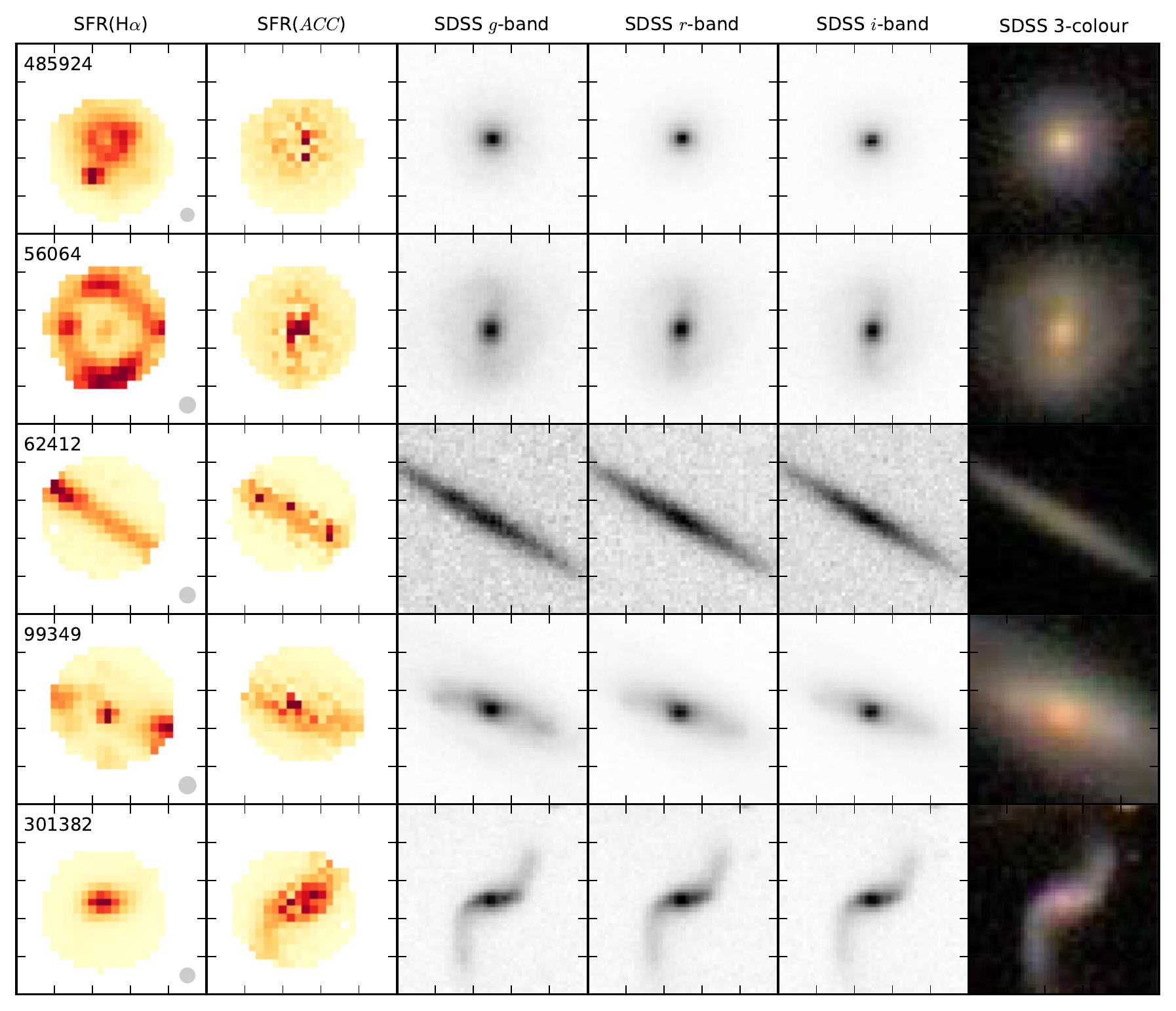}}
\caption{The SFR(\Ha) map, SFR($ACC$) map, SDSS $g,r,i$-band images and the SDSS $3$-colour image for $5$ galaxies (a galaxy per row). Both SFR maps have been normalised to the maximum of each map respectively. The $g$-band PSF size for the SAMI data is shown by a grey circle in the lower right of the SFR(\Ha) maps. The SFR maps are made using SAMI cubes that have been binned to have $1$~arcsec spaxels (native spaxel size is $0.5$~arcsec). All maps and images are $25\times25$~arcsec in size, and are orientated such that North is up and East is left. The SAMI galaxy ID is provided in the upper left of the SFR(\Ha) maps for reference in the text. These galaxies have been selected to highlight differences between the SFR maps. Only a few galaxies not represented here have smooth SFR maps that closely match each other. \vspace{-0.2cm}}
\label{fig:SFR_maps}
\end{figure*}

With the $ACC$ constructed, for each galaxy it is possible to compare the SFR(\Ha) map to its SFR($ACC$) map. A galaxy's SFR($ACC$) map is made by locating the $u^s,g^s,r^s$ colours of a spaxel on the $u^s$-- $g^s$, $g^s$-- $r^s$ grid of the $ACC$ and multiplying the SFR/$r^s$-band luminosity likelihood distribution of that cell by the spaxel's $r^s$-band luminosity. The SFR is taken as the median of the likelihood distribution. Regardless of the spatial distribution of the SFR(\Ha), the SFR($ACC$) followed a smooth distribution tracing out the optical continuum. This discontinuity is enhanced for more complex SFR(\Ha) distributions (see Figure \ref{fig:SFR_maps} for a selection of these maps). 

\begin{figure*}
\centering
\centerline{\includegraphics[width=15.575cm]{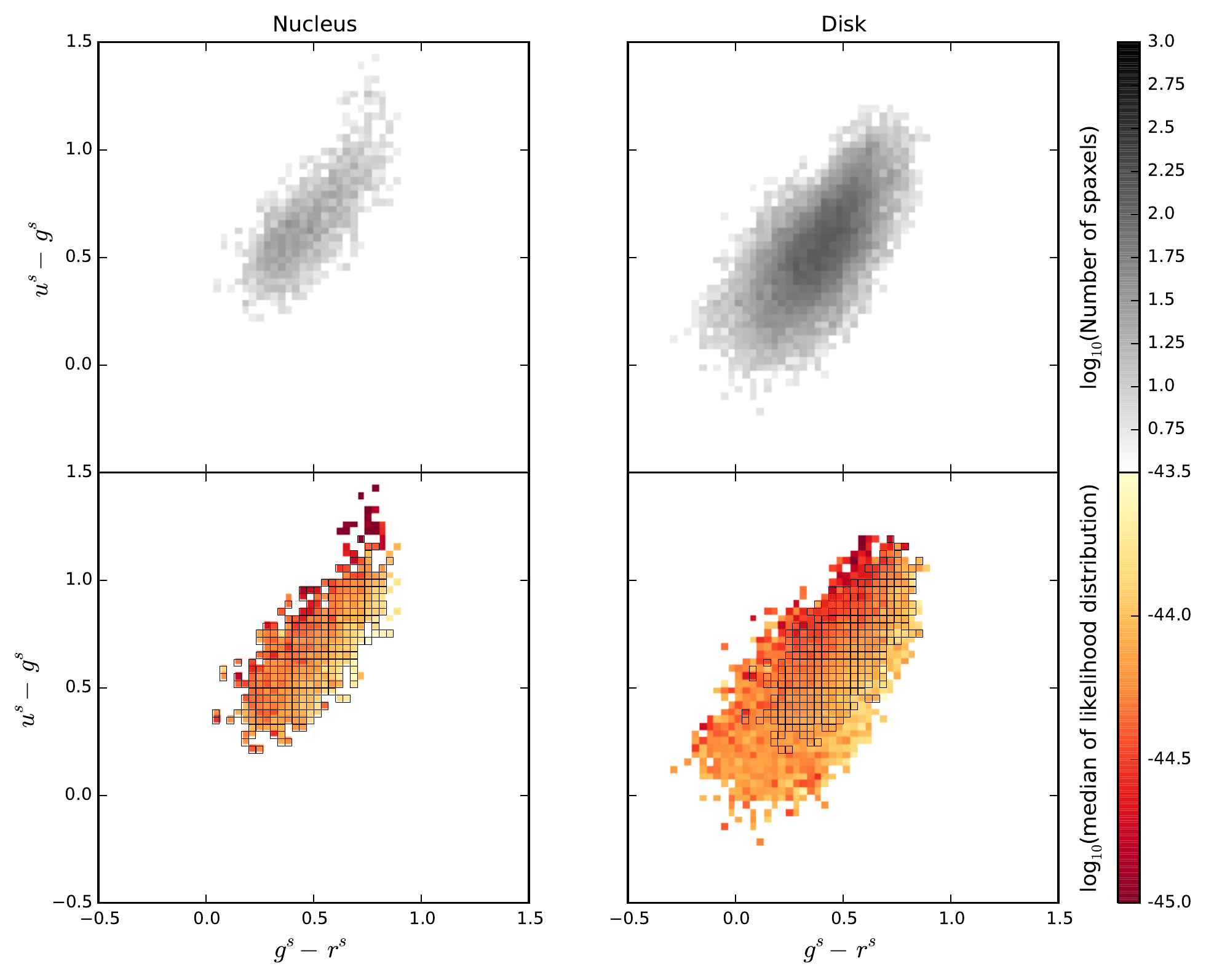}}
\caption{Two aperture correction cubes ($ACC$); one built from only spaxels in the central $3$~arcsec of the galaxy (Nucleus, left column), and the other built from spaxels outside the central $3$~arcsec (Disk, right column). The diagrams follow the same description as Figure \ref{fig:B04_cubes}. Cells common between both $ACC$s are outlined in the lower panels. The disk $ACC$ covers a larger range of colours (although, missing the reddest of spaxels with high $u^s$-- $g^s$).}
\label{fig:B04_ND}
\end{figure*}

The second assumption from B04 that we can examine is that an aperture correction cube built from nuclear spectra can be representative of the SFR in the disk of a galaxy. Here we proceed to build two $ACC$s in the same fashion as before, though this time with: (1) only spaxels contained in the central $3$~arcsec diameter of the galaxy (nuclear); (2) only spaxels outside the central $3$~arcsec (disk). These $ACC$s can be seen in Figure \ref{fig:B04_ND}. The most obvious difference is that the disk $ACC$ spans a larger range of colours, but doesn't probe as far into $u^s$-- $g^s$ as the nucleus $ACC$. This is expected as the nuclear region of galaxies tend to have redder colours due to the presence of older stars. Another difference arises in the likelihood distributions, with the medians of the nuclear $ACC$ changing more rapidly than the disk $ACC$ in the $u^s$-- $g^s$, $g^s$-- $r^s$ plane. This difference can be seen more easily in Figure \ref{fig:B04_ND_MvBD}, where the data points are $u^s$-- $g^s$, $g^s$-- $r^s$ cells that overlap between the nuclear $ACC$ and the disk $ACC$. A positive correlation is found between the difference of the likelihood distribution medians for each $ACC$ and the respective median Balmer decrements for a given $u^s$-- $g^s$, $g^s$-- $r^s$ cell. When the nuclear spectra under-estimate the Balmer decrement for the disk, the SFR derived from an aperture correction cube built from only nuclear spectra is over-estimated. Inversely, when the nuclear spectra over-estimate the Balmer decrement for the disk, the SFR is under-estimated. The histogram of the differences has a median value of $0.04$~dex (under-estimation of SFR) and a $1\sigma$~scatter of $0.16$~dex. Whilst examining the effect of this correlation on the B04 against SAMI SFRs in Figure \ref{fig:SAMI_H03_B04}, we also found a positive correlation between the total SFR of a galaxy and the ratio of median Balmer decrement for spaxels within a $3$~arcsec diameter aperture (nuclear) to the median Balmer decrement for remaining spaxels (disk) (see Figure \ref{fig:B04_ND_B04vBD}).The spaxels that contributed to both $ACC$s occupied the same star forming sequence on a log$_{10}$(\OHb) vs log$_{10}$(\NHa) BPT diagram, ruling out contamination of other ionisation sources to the nuclear spectra.

\vspace{0.0cm}
\section{Discussion}
\vspace{0.2cm}
\subsection{H03 method (GAMA)}
The attempt of disentangling random and systematic errors from SFR comparison plots, such as Figure \ref{fig:SAMI_H03_B04}, can prove to be difficult, if not impossible. When comparing H03 SFRs and SAMI SFRs we find a near $1$:$1$ trend (gradient of $0.91\pm0.05$ with a $1\sigma$~scatter of $0.22$~dex). Deviation from $1$:$1$ happens for low-SF galaxies, with H03 over-predicting the SFR. Studies of dwarf galaxies in the local universe (which occupy the low star forming end of the H03 against SAMI SFRs in Figure \ref{fig:SAMI_H03_B04}) have been shown to exhibit bursty star formation, in addition to an underlying ageing population \citep{2003ApJS..147...29G,2014MNRAS.445.1104R}. An order of magnitude in the difference of timescales leads to an $r$-band continuum level over-representing the instantaneous (\Ha) star formation. The large scatter can be understood with the analysis of the SFR curves-of-growth (see Figure \ref{fig:H03_cogs}), where galaxies with a small aperture ($d/R_{\rm 2P}<0.4$) have an uncertainty on their aperture corrected SFR~$\approx0.25$~dex. The high dispersion of the H03 curves-of-growth at small apertures can also be seen in the work of \citet{2013A&A...553L...7I} who at small apertures find large dispersions in the EW(\Ha) and Balmer decrement profiles of $107$ {\small \sc CALIFA} galaxies with SFRs $\gtrsim1$~\Moy. Finding no correlation between the H03 curves-of-growth and another global galaxy parameter results in an interpretation that the H03 error \,(Table \ref{tab:H03_cogs_fits}) is random. The trend in the medians of these distributions, however, suggests that H03 are systematically over-estimating their SFRs by up to $0.1$~dex for galaxies with the smallest apertures ($d/R_{\rm 2P}<0.2$). The analytical expressions of these error distributions (Table \ref{tab:H03_cogs_fits}) can be used, together with measurement errors of the EW(\Ha) and Balmer decrement, to obtain formal errors on the H03 SFRs.

The random nature of the H03 error should only be considered to be valid with an unbiased sample selection. Adopting a sample selection that could bias the EW(\Ha) curves-of-growth will also introduce biases in the H03 SFRs. Such science can include investigation into the trends in SFR for merging galaxies, as star formation is seen to be more centrally concentrated in these systems, which will lead to an over-estimation of the H03 SFRs \citep[][Bloom et al. \emph{in prep}]{2015MNRAS.448.1107M}. Galaxies with centrally concentrated star formation are also more likely to be found in higher density environments \citep[][Schaefer et al. \emph{in prep}]{2006AJ....131..716K, 2012A&A...544A.101C}, where the H03 SFRs would also become over-estimated, although in this work we found no statistically significant correlation between the H03 SFR curves-of-growth and environment. 

For GAMA DR2 galaxies with $z<0.1$, the median $d/R_{\rm 2P}\approx0.3$ and H03 $1\sigma$ error $\approx0.18$~dex. Results such as the \Ha\ luminosity function presented by \citet{2013MNRAS.433.2764G} will be affected by this error. The over-estimation bias of up to $0.1$~dex for apertures with $d/R_{\rm 2P}<0.2$ can lead to a steeper turn off in the shape of the \Ha\ luminosity function at the high luminosity end (Gunawardhana, \emph{private communication}). This section of the \Ha\ luminosity function is where you tend to find larger galaxies (higher \Ha\ luminosity), so the $d/R_{\rm 2P}$ aperture size would be smaller.

\begin{figure}
\centering
\centerline{\includegraphics[width=\linewidth]{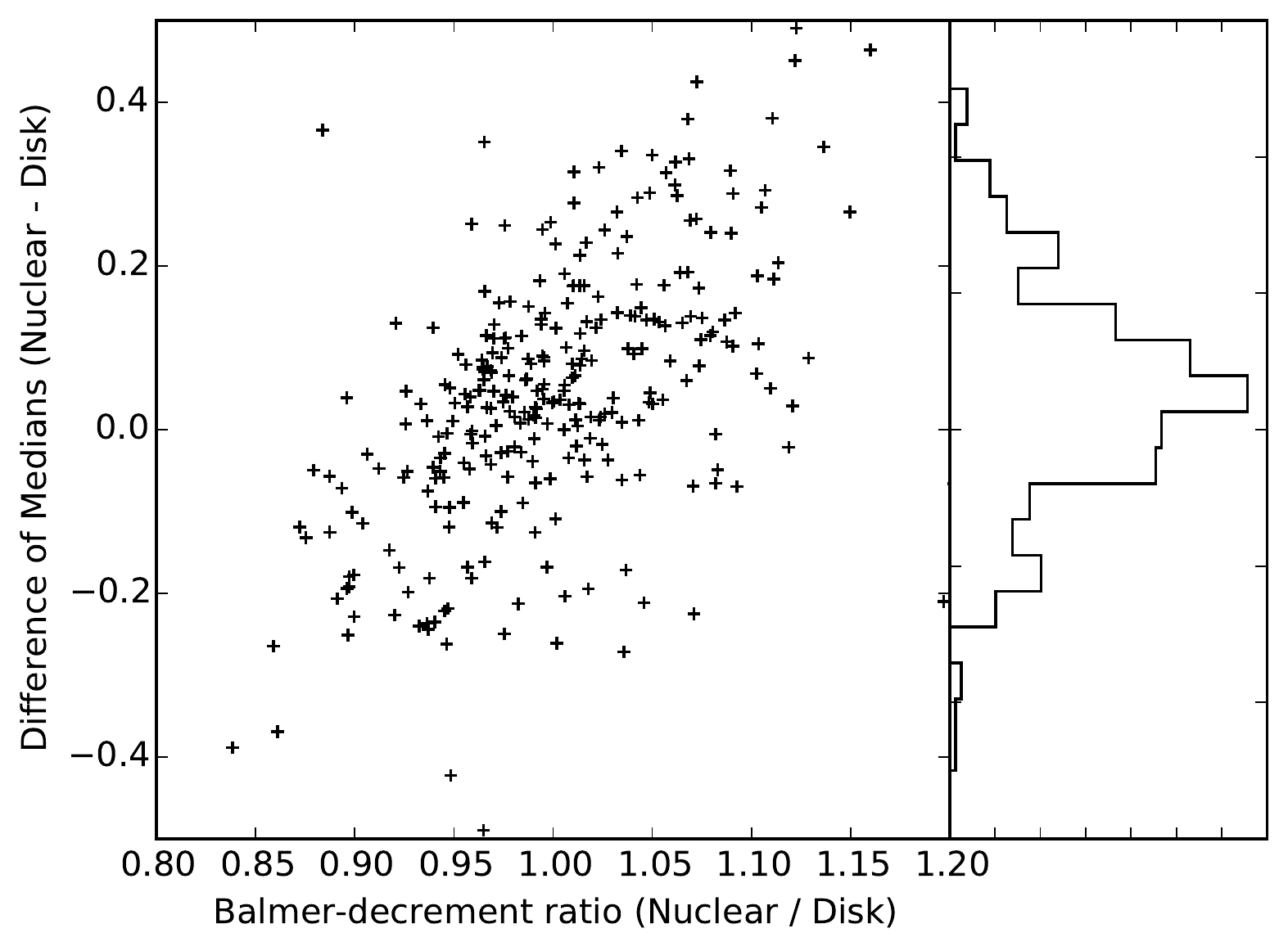}}
\caption{The difference of the median of the likelihood distributions (in dex) against the ratio of the median Balmer decrement for each common $u^s$-- $g^s$, $g^s$-- $r^s$ cell in the aperture correction cubes (nuclear $ACC$ and the disk $ACC$, see Figure \ref{fig:B04_ND}). The Spearman's rank correlation coefficient is $0.561$ with a p-value of $1.68\times10^{-23}$. The histogram shows the distribution of the differences, which has a median of $0.04$~dex and a $1\sigma$-error of $0.16$~dex. For positive difference the nuclear $ACC$ under-predicts the SFR found from the disk $ACC$, and vice versa.}
\label{fig:B04_ND_MvBD}
\end{figure} 

\begin{figure}
\centering
\centerline{\includegraphics[width=\linewidth]{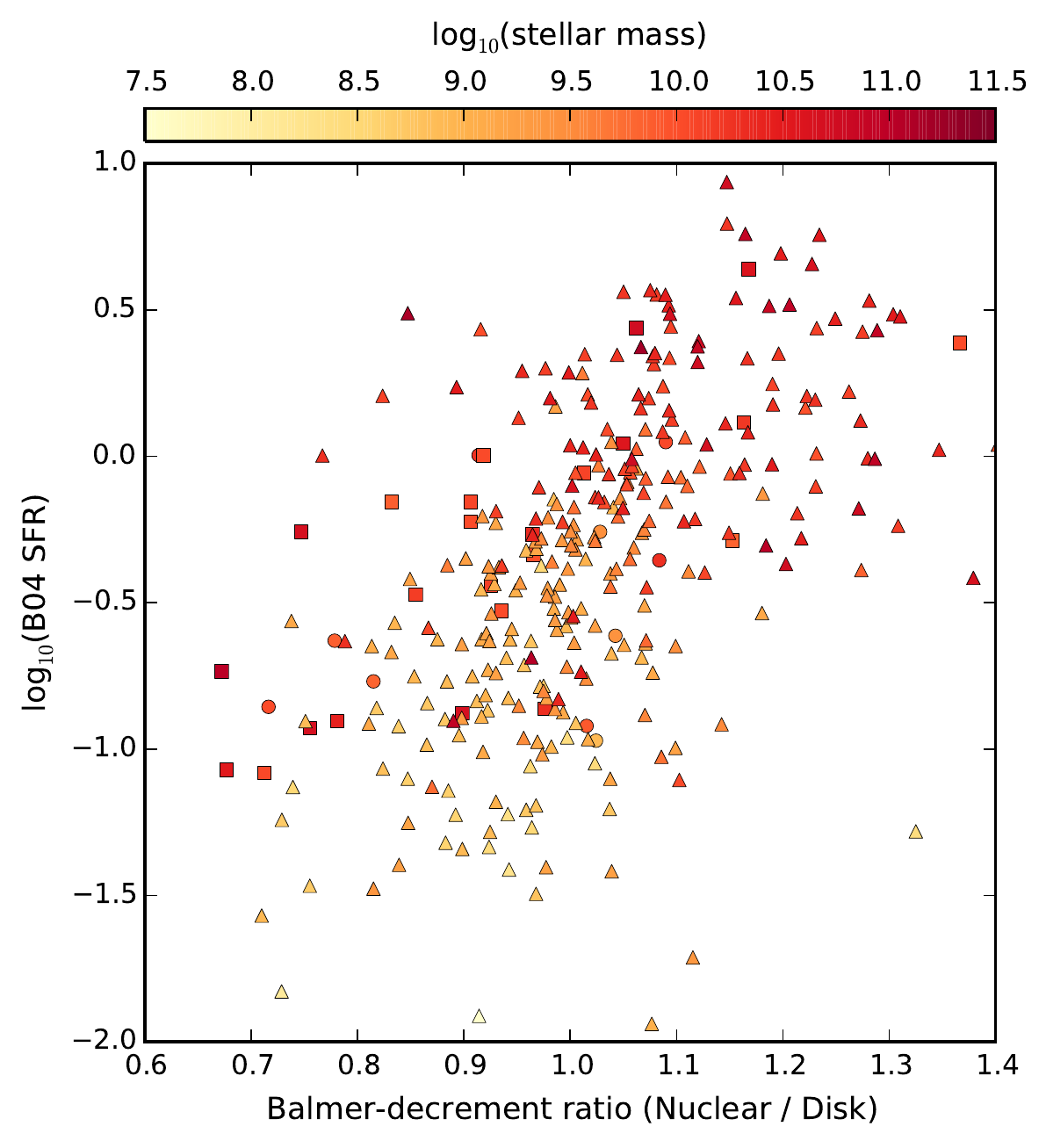}}
\caption{The log$_{10}$(B04 SFRs) for $337$ star forming (SF) SAMI galaxies against the ratio of the median Balmer decrement for spaxels within a $3$~arcsec aperture (nuclear) to the median Balmer decrement for spaxels outside a $3$~arcsec aperture (disk). Square, triangle and circle markers represent early-type, late-type and unclassified morphologies respectively, and are coloured with respect to each galaxy's stellar mass. The Spearman's rank correlation coefficient is $0.595$ with a p-value of $1.46\times10^{-26}$. Galaxies with a higher SFR (or higher stellar mass) tend to have more dust in their disk compared to their nucleus.}
\label{fig:B04_ND_B04vBD}
\end{figure}

\vspace{-0.4cm}
\subsection{B04 method (SDSS)}
Understanding the slope of the B04 SFRs and SAMI SFRs from Figure \ref{fig:SAMI_H03_B04} required the creation of an aperture correction cube based on SAMI data ($ACC$) to discover how well broadband colours could trace the \Ha\ based star formation. Figure \ref{fig:SFR_maps} shows the SFR distributions of a selection of galaxies with SFR maps measured from \Ha\ and the $ACC$. Here we find that the SFR($ACC$) traces out the broadband light of the galaxy even when the SFR(\Ha) is more clumpy. Clear examples of this from Figure \ref{fig:SFR_maps} are SAMI IDs {\small 485924} and {\small 56064}. {\small 485924} has an off-centre starburst which is not detected in the broadband imaging or in the SFR($ACC$). {\small 56064} appears to have most of its star formation in the disk, but the SFR($ACC$) predicts more star formation in the nucleus. Similarly to the analysis of H03 (see Figure \ref{fig:H03_EW_BD}), only $\sim1/3$ of our galaxies exhibit a smooth distribution of SFR(\Ha) that closely matches the distribution of SFR($ACC$). Making SFR maps is not what the B04 aperture correction was intended for, although it highlights the need for IFS surveys of many thousands of galaxies.

In an aperture correction cube there is a degeneracy between stellar age, metallicity and dust that broadens the likelihood distributions, though the underlying issue is assuming the optical continuum (timescales of $>100$~Myr) can trace SFR on timescales $<10$~Myr. The effect of not being sensitive to starbursts can be one explanation to the under-estimation in B04 SFRs for galaxies with a high SFR in Figure \ref{fig:SAMI_H03_B04}. Although these galaxies are more likely to have bluer colours, they also have a tendency to have more prominent starbursts, resulting in the under-estimation of B04 SFR. This under-estimation has also been seen by Green et al. \emph{in prep}, who compare the B04 SFRs with total \Ha\ SFRs from IFU data of $67$ galaxies with SFRs of $1$ to $100$ \Moy. \citet{2007ApJS..173..267S} found that for galaxies with B04 SFR of $1$ to $30$ \Moy, the SFRs matched with SFRs derived from $FUV,NUV,u,g,r,i,z$ broadband measurements. This match is expected because the two star formation measures probe similar timescales.

Figure \ref{fig:B04_ND} shows a difference in the medians of the likelihood distributions of the nuclear $ACC$ and disk $ACC$, meaning that the assumption in the B04 method that an $ACC$ built from nuclear spectra can be representative of the galaxy as a whole has underlying errors. Investigating this difference further, we find a positive correlation between the medians of the likelihood distributions and the medians of the Balmer decrements for $u^s$-- $g^s$, $g^s$-- $r^s$ cells that are common between both $ACC$s (see Figure \ref{fig:B04_ND_MvBD}). This is a probe into the degeneracy of dust in an aperture correction cube. Galaxies with strong increasing or decreasing dust gradients will have over- or under-predicted B04 SFRs respectively. The dust gradient of a galaxy correlates with its total SFR (or stellar mass, see Figure \ref{fig:B04_ND_B04vBD}), such that high star forming galaxies (or greater stellar mass) tend to have decreasing dust gradients, and low star forming galaxies have increasing dust gradients. \citet{2013A&A...553L...7I} also find that galaxies with SFRs $\gtrsim 1$~\Moy\ have a decreasing dust gradient. This correlation might explain the slope in the B04 against SAMI SFRs from Figure \ref{fig:SAMI_H03_B04}. The B04 SFRs for high star forming galaxies are under-predicted compared to SAMI. This under-representation arises when deriving SFRs from an aperture correction cube that is built only using nuclear spectra. B04 also over-predict the SFR for low star forming galaxies for the same reason. 

The B04 slope from Figure \ref{fig:SAMI_H03_B04} requires a correction term based on these correlations. However, due to the difference in broadband filters used in B04 and this work to create the aperture correction cubes, we are unable to provide this correction. To find the true correction term, nuclear and disk aperture correction cubes would need to be built from IFS data of $\sim10^3$ galaxies that spectrally cover the $g,r,i$ filters. This will be possible with {\small MaNGA}{\sc} or {\small HECTOR}{\sc} \citep{2012SPIE.8446E..53L,2015IAUS..309...21B}. With these larger surveys, it will also be possible to investigate any biases that arise in the B04 method due to stellar age and metallicity. 

In the age of multi-wavelength surveys such as {\small GAMA}{\sc} \citep{2015MNRAS.452.2087L}, analogous aperture correction cubes can be built from many different SFR indicators, and comparisons made to further identify possible biases. Any tracer of SFR can be used in the construction of an aperture correction cube, though the cube would be sensitive to different timescales of star formation. 

\vspace{-0.4cm}
\section{Conclusions}
We have used integral-field spectroscopy of $1212$ galaxies from the SAMI Galaxy Survey to probe the assumptions that underpin the \Ha\ star formation rate aperture correction methods of \citet[][H03]{2003ApJ...599..971H} and \citet[][B04]{2004MNRAS.351.1151B}. We summarise the findings of this work:
\vspace{-0.2cm}
\begin{enumerate}
	\item When comparing total star formation rates (SFRs) from the H03 and B04 aperture corrections with integrated \Ha\ SFRs from SAMI data, both H03 and B04 have trends that deviate from $1$:$1$. The gradient and scatter for H03/SAMI are $0.91\pm0.05$ and $0.22$~dex, and for B04/SAMI are $0.85\pm0.03$ and $0.15$~dex. 
	\item Only $\approx1/3$ of our galaxies follow H03's assumption that the EW(\Ha) and Balmer decrement curves-of-growth remain flat. For the sample considered here, the likelihood of increasing or decreasing curves-of-growth is the same. Our empirically derived, analytical expression of the error on and correction for this assumption can be found in Table \ref{tab:H03_cogs_fits}. Using it, the median GAMA DR2 galaxy with redshift $z<0.1$ has an H03 SFR $1\sigma$~error of $0.18$~dex (not inclusive of measurement errors on EW(\Ha) and Balmer decrement).
	\item Investigations into the B04 method showed that although this method includes a dependance on optical colours, and is therefore more sensitive to younger, hotter stars, the SFRs found can still be insensitive to starbursts (instantaneous star formation). This is because the \Ha\ emission and optical continuum probe two different timescales ($<10$ and $>100$ Myr respectively).
	\item We compared two aperture corrections similar to B04 from SAMI data, built from spectra of the nuclear regions of galaxies and separately from spectra beyond. We found B04's assumption that nuclear spectra can be representative of the rest of the galaxy to be biased due to a difference in the nuclear and disk dust corrections. 
	\item We find that the dust gradient and total SFR of a galaxy are correlated such that galaxies with a high SFR require a smaller dust correction in their disk compared to their nucleus. This results in an under-estimation of the total SFR when using a B04 aperture correction method built only from nuclear spectra. This bias is also seen in low star forming galaxies requiring a larger dust correction in their disk compared to their nucleus, resulting in an over-estimation in SFR. The slope found when comparing total SFRs of star forming galaxies from B04 and SAMI can be explained by these correlations.
	\item Measuring the magnitude of the bias in the B04 aperture correction requires further investigation using IFS data that covers the same wavebands (e.~g.~{\small MaNGA} or {\small HECTOR}).
	\item A sample selection that prefers galaxies with concentrated or extended star formation will bias the H03 SFRs to be over- or under-estimated respectively. Whereas, a sample selection that prefers galaxies with high or low star formation will bias the B04 SFRs. Choosing which aperture correction is suitable to minimise any potential bias will depend on the data sample in question. 
\end{enumerate}
\vspace{-0.2cm}
So, "Can we trust aperture corrections to predict star formation?". Yes, but only for large ($\gtrsim10^3$) unbiased samples of galaxies, and as long as the conclusions can have accuracies of $\sim0.2$~dex in SFR. At this level of uncertainty, there are two main cases of preference between the \citet[][H03]{2003ApJ...599..971H} and \citet[][B04]{2004MNRAS.351.1151B} aperture correction methods: (a) the inclusion of galaxies classified outside of the star formation main sequence in BPT diagnostics is only possible in the B04 method; (b) the H03 method has lower systematic biases over a large dynamic range in SFR for complete data samples. 

\vspace{-0.4cm}
\section{Acknowledgments}

The SAMI Galaxy Survey is based on observation made at the Anglo-Australian Telescope. The Sydney-AAO Multi-object Integral field spectrograph (SAMI) was developed jointly by the University of Sydney and the Australian Astronomical Observatory. The SAMI input catalogue is based on data taken from the Sloan Digital Sky Survey, the GAMA Survey and the VST ATLAS Survey. The SAMI Galaxy Survey is funded by the Australian Research Council Centre of Excellence for All-sky Astrophysics (CAASTRO), through project number CE110001020, and other participating institutions. The SAMI Galaxy Survey website is http://sami-survey.org/.

The ARC Centre of Excellence for All-sky Astrophysics (CAASTRO) is a collaboration between The University of Sydney, The Australian National University, The University of Melbourne, Swinburne University of Technology, The University of Queensland, The University of Western Australia and Curtin University, the latter two participating together as the International Centre for Radio Astronomy Research (ICRAR). CAASTRO is funded under the Australian Research Council (ARC) Centre of Excellence program, with additional funding from the seven participating universities and from the NSW State Government's Science Leveraging Fund.

Funding for SDSS-III has been provided by the Alfred P. Sloan Foundation, the Participating Institutions, the National Science Foundation, and the U.S. Department of Energy Office of Science. The SDSS-III web site is http://www.sdss3.org/.

GAMA is a joint European-Australasian project based around a spectroscopic campaign using the Anglo-Australian Telescope. The GAMA website is http://www.gama-survey.org/.

SMC acknowledges support from an ARC Future fellowship (FT100100457). JTA acknowledges the award of an ARC Super Science Fellowship (FS110200013). MSO acknowledges the funding support from the Australian Research Council through a Future Fellowship (FT140100255) MLPG acknowledges support from a European Research Council grant (DEGAS-259586). LC acknowledges support under the Australian Research Council's Discovery Project funding scheme (DP130100664).

\vspace{-0.2cm}



\bibliographystyle{mn2e}
\bibliography{_papers}



\end{document}